\documentclass[zpreprint,zbstnp]{zeus_paper}
\usepackage[english]{babel}
\usepackage{multicol}
\newcommand{\ZcoosysA}{%
The ZEUS coordinate system is a right-handed Cartesian system, with the $Z$
axis pointing in the proton beam direction, referred to as the ``forward
direction'', and the $X$ axis pointing left towards the center of HERA.
The coordinate origin is at the nominal interaction point.\xspace}

\newcommand{\ZcoosysfnA}{\footnote{\ZcoosysA}}

\newcommand{\Zdetdesc}{%
A detailed description of the ZEUS detector can be found 
elsewhere~\cite{zeus:1993:bluebook}. A brief outline of the 
components that are most relevant for this analysis is given
below.\xspace}
\newcommand{\Zctddesc}[1]{%
Charged particles are tracked in the central tracking detector (CTD)~\citeCTD,
which operates in a magnetic field of $1.43\Tesla$ provided by a thin 
superconducting coil. 
The CTD consists of 72~cylindrical drift chamber 
layers, organized in 9~superlayers covering the polar-angle\ZcoosysfnA ~region 
\mbox{$15^\circ<\theta<164^\circ$}. The transverse-momentum resolution for
full-length tracks is $\sigma(p_T)/p_T=0.0058p_T\oplus0.0065\oplus0.0014/p_T$,
with $p_T$ in $\Gev$.\xspace}
\newcommand{\Zcaldesc}{%

The high-resolution uranium--scintillator calorimeter (CAL)~\citeCAL consists 
of three parts: the forward, %(FCAL) 
the barrel %(BCAL) 
and the rear (RCAL) calorimeters. 
Each part is subdivided transversely into towers and
longitudinally into one electromagnetic section %(EMC) 
and either one %(in RCAL)
or two %(in BCAL and FCAL) 
hadronic sections. %(HAC). 
The smallest subdivision of
the calorimeter is called a cell.  The CAL energy resolutions, as measured under
test-beam conditions, are $\sigma(E)/E=0.18/\sqrt{E}$ for electrons and
$\sigma(E)/E=0.35/\sqrt{E}$ for hadrons, with $E$ in $\Gev$.
}%
\chardef\usc=95
\chardef\til=126
\catcode`\@=11 % @ signs are now treated as letters
\DeclareRobustCommand\xdotspace{\futurelet\@let@token\@xdotspace}
\def\@xdotspace{%
  \ifx\@let@token.\else
  \ifx\@let@token\bgroup.\else
  \ifx\@let@token\egroup.\else
  \ifx\@let@token\/.\else
  \ifx\@let@token\ .\else
  \ifx\@let@token~.\else
  \ifx\@let@token!.\else
  \ifx\@let@token,.\else
  \ifx\@let@token:.\else
  \ifx\@let@token;.\else
  \ifx\@let@token?.\else
  \ifx\@let@token/.\else
  \ifx\@let@token'.\else
  \ifx\@let@token).\else
  \ifx\@let@token-.\else
  \ifx\@let@token\@xobeysp.\else
  \ifx\@let@token\space.\else
  \ifx\@let@token\@sptoken.\else
   .\space
   \fi\fi\fi\fi\fi\fi\fi\fi\fi\fi\fi\fi\fi\fi\fi\fi\fi\fi}
\catcode`\@=12 % @ signs are no longer letters

\newcommand{\stru}[2]{%
   \relax\ifmmode\hbox{\vrule height#1 depth#2 width0pt}%
   \else\vrule height#1 depth#2 width0pt\fi}
\newcommand{\Ronum}[1]{\uppercase\expandafter{\romannumeral#1}}
\newcommand{\ronum}[1]{\expandafter{\romannumeral#1}}
\DeclareRobustCommand{\LaTeXZ}{%
  \LaTeX\kern-.05em4\kern-.1em
  {\raisebox{-0.2ex}{$\scriptstyle\text{ZEUS}$}}\xspace}
\DeclareMathAlphabet{\mathbf}{OT1}{cmr}{bx}{sl}
\newcommand{\eVdist}{\kern-0.06667em}

\newcommand{\Gev}{{\text{Ge}\eVdist\text{V\/}}}

\newcommand{\mev}{{\,\text{Me}\eVdist\text{V\/}}}
\newcommand{\gev}{{\,\text{Ge}\eVdist\text{V\/}}}

\newcommand{\pbi}{\,\text{pb}^{-1}}

\newcommand{\cm}{\,\text{cm}}

\newcommand{\Tesla}{\,\text{T}}

\newcommand{\slashfrac}[2]{%
  \raisebox{0.5ex}{\ensuremath #1}\kern-0.12em/\kern-0.08em
  \raisebox{-.8ex}{\ensuremath #2}}
\newcommand{\sqr}[3]{%
    {\vcenter{\hrule height.#3ex\hbox{\vrule width.#2ex height#1ex
     \kern#1ex\vrule width.#3ex}\hrule height.#2ex}}}

\catcode`\@=11 % @ signs are now treated as letters
\newcommand{\parenbar}{\mathpalette\p@renb@r}
\def\p@renb@r#1#2{\vbox{%
  \ifx#1\scriptscriptstyle \dimen@.7em\dimen@ii.2em\else
  \ifx#1\scriptstyle \dimen@.8em\dimen@ii.25em\else
  \dimen@1em\dimen@ii.4em\fi\fi \offinterlineskip
  \ialign{\hfill##\hfill\cr
    \vbox{\hrule width\dimen@ii}\cr
    \noalign{\vskip-.3ex}%
    \hbox to\dimen@{$\mathchar300\hfil\mathchar301$}\cr
    \noalign{\vskip-.3ex}%
    $#1#2$\cr}}}
\catcode`\@=12 % @ signs are no longer letters

\newcommand{\IP}{{\rm I$\kern-0.01667em$P}\xspace}

\mathchardef\qsm=63
\mathchardef\pls=43
\mathchardef\mns=512
\mathchardef\plm=518
\mathchardef\eql=61
\mathchardef\smallleft=300
\mathchardef\smallright=301
\mathchardef\les=316
\mathchardef\gre=318
\mathchardef\leq=532
\mathchardef\grq=533
\catcode`\@=11 % @ signs are now treated as letters
\newcounter{pict@width}
\newcounter{pict@height}
\newlength{\pict@scale}
\newcommand{\psfigadd}[4]{%
\setcounter{pict@width}{1*\ratio{#2+\pict@scale/2}{\pict@scale}}
\setcounter{pict@height}{1*\ratio{#3+\pict@scale/2}{\pict@scale}}
\setlength{\unitlength}{\pict@scale}
\hbox to #2{\hspace{-\fill}\begin{picture}(\thepict@width,\thepict@height)
\put(0,0){\psfig{figure=#1,width=#2,height=#3,clip=}}
\SetScale{0.283466457}
\SetWidth{1.763889}
{#4}
\end{picture}}
}
\newcounter{pict@widthfst}
\newcounter{pict@widthscd}
\newcounter{pict@widthtot}
\newcommand{\psfigaddtwo}[7]{%
\setcounter{pict@widthfst}{1*\ratio{#2+\pict@scale/2}{\pict@scale}}
\setcounter{pict@widthscd}{1*\ratio{#2+#4+\pict@scale/2}{\pict@scale}}
\setcounter{pict@widthtot}{1*\ratio{#2+#4+#6+\pict@scale/2}{\pict@scale}}
\setcounter{pict@height}{1*\ratio{#3+\pict@scale/2}{\pict@scale}}
\setlength{\unitlength}{\pict@scale}
\hbox{\hspace{-\fill}\begin{picture}(\thepict@widthtot,\thepict@height)
\put(0,0){\psfig{figure=#1,width=#2,height=#3,clip=}}
\put(\thepict@widthscd,0){\psfig{figure=#5,width=#6,height=#3,clip=}}
\SetScale{0.283466457}
\SetWidth{1.763889}
{#7}
\end{picture}}
}
\newcommand{\psfigror}[4]{%
\setcounter{pict@width}{1*\ratio{#2+\pict@scale/2}{\pict@scale}}
\setcounter{pict@height}{1*\ratio{#3+\pict@scale/2}{\pict@scale}}
\setlength{\unitlength}{\pict@scale}
\hbox{\begin{picture}(\thepict@width,\thepict@height)
\put(0,\thepict@height){\psfig{figure=#1,width=#3,height=#2,clip=,angle=270}}
\SetScale{0.283466457}
\SetWidth{1.763889}
{#4}
\end{picture}}
}
\newcommand{\psfigrol}[4]{%
\setcounter{pict@width}{1*\ratio{#2+\pict@scale/2}{\pict@scale}}
\setcounter{pict@height}{1*\ratio{#3+\pict@scale/2}{\pict@scale}}
\setlength{\unitlength}{\pict@scale}
\hbox{\begin{picture}(\thepict@width,\thepict@height)
\put(0,0){\psfig{figure=#1,width=#3,height=#2,clip=,angle=90}}
\SetScale{0.283466457}
\SetWidth{1.763889}
{#4}
\end{picture}}
}
\catcode`\@=12 % @ signs are no longer letters
\newlength\listtextwidth

\catcode`\@=11 % @ signs are now treated as letters
\newlength{\@tabfninsert}
\newlength{\@tabfnwidth}
\newcommand{\tabfootnote}[2]{%
  \setlength{\@tabfninsert}{0.8em}
  \setlength{\@tabfnwidth}{\textwidth}
  \addtolength{\@tabfnwidth}{-\@tabfninsert}
  \addtolength{\@tabfnwidth}{-0.4em}
  \noindent\makebox[\@tabfninsert][r]{\footnotesize$^{#1}$\hfil}\hfill%
  \parbox[t]{\@tabfnwidth}{\footnotesize #2\hfill}}
\catcode`\@=12 % @ signs are no longer letters

\newcommand{\ZdedxA}{%                                                        
To estimate the energy loss, $dE/dx$, of tracks, the truncated mean of the 
sense-wire pulse-heights was recorded for each track, discarding the $10\%$
lowest and up to the $30\%$ highest pulses \mcite{pl:b481:213,epj:c18:625,dedx}.
The measured $dE/dx$ values were normalised to the $dE/dx$ peak position for 
tracks with momenta $0.3<p<0.4$ \gev, the region of minimum ionisation for pions. 
Henceforth, $dE/dx$ is quoted in units of minimum ionising particles (mips).
The resolution of the $dE/dx$ measurement for full-length tracks is about
$9\%$. The tracking system was used to establish the primary and secondary 
vertices.\xspace} 
 
\xspace                                                                       
\def\citeCTD{{\cite{%                                                           
nim:a279:290,*npps:b32:181,*nim:a338:254%                                       
}}\xspace}                                                                      
\def\citeCAL{{\cite{%                                                           
nim:a309:77,*nim:a309:101,*nim:a321:356,*nim:a336:23%                           
}}\xspace}

\begin{document}
\prepnum{DESY 07-069}                                                           
\date{}
%------------------------------------------------------------------------------
%       Title sheet
%------------------------------------------------------------------------------
\hyphenation{stu-died za-wiej-ski}                                                          
\title{                                                                         
Bose-Einstein Correlations of Charged and\\                                     
Neutral Kaons in Deep Inelastic Scattering\\                                    
at HERA                                                                         
}                                                                               
\author{ZEUS Collaboration}                                                     
\abstract{                                                                      
Bose-Einstein correlations of charged and neutral kaons                         
have been measured in $e^{\pm}p$ deep inelastic scattering with                 
an integrated luminosity of $121\pbi$ using the ZEUS detector at HERA.          
The two-particle correlation function was studied as a function of              
the four-momentum difference of the kaon pairs,                                 
$Q_{12} = \sqrt{-(\textrm{p}_1-\textrm{p}_2)^2}$,                               
assuming a Gaussian shape for the particle source.                              
The values of the radius of the production volume, {\it r}, and of the          
correlation strength, $\lambda$, were obtained for both neutral and             
charged kaons. The radii for charged and neutral kaons are similar and          
are consistent with those obtained at LEP.                                      
}                                                                               
\makezeustitle  
%------------------------------------------------------------------------------
%       ZEUS Authors 
%------------------------------------------------------------------------------
\def\3{\ss}                                                                                        
\pagenumbering{Roman}                                                                              
\begin{center}                                                                                     
{                      \Large  The ZEUS Collaboration              }                               
\end{center}                                                                                       
  S.~Chekanov$^{   1}$,                                                                            
  M.~Derrick,                                                                                      
  S.~Magill,                                                                                       
  B.~Musgrave,                                                                                     
  D.~Nicholass$^{   2}$,                                                                           
  \mbox{J.~Repond},                                                                                
  R.~Yoshida\\                                                                                     
 {\it Argonne National Laboratory, Argonne, Illinois 60439-4815}, USA~$^{n}$                       
\par \filbreak                                                                                     
  M.C.K.~Mattingly \\                                                                              
 {\it Andrews University, Berrien Springs, Michigan 49104-0380}, USA                               
\par \filbreak                                                                                     
  M.~Jechow, N.~Pavel~$^{\dagger}$, A.G.~Yag\"ues Molina \\                                        
  {\it Institut f\"ur Physik der Humboldt-Universit\"at zu Berlin,                                 
           Berlin, Germany}
% ...

\newpage                                                                                           
$^{\    1}$ supported by DESY, Germany \\                                                          
$^{\    2}$ also affiliated with University College London, UK \\                                  
% ...
$^{\dagger}$ deceased \\                                                                           

\nopagebreak
\begin{tabular}[!h]{rp{14cm}}                                                                       
$^{a}$ &  supported by the Natural Sciences and Engineering Research Council of Canada (NSERC) \\  
$^{b}$ &  supported by the German Federal Ministry for Education and Research (BMBF), under        
          contract numbers HZ1GUA 2, HZ1GUB 0, HZ1PDA 5, HZ1VFA 5\\                                
$^{c}$ &  supported in part by the MINERVA Gesellschaft f\"ur Forschung GmbH, the Israel Science   
          Foundation (grant no. 293/02-11.2) and the U.S.-Israel Binational Science Foundation \\  
$^{d}$ &  supported by the German-Israeli Foundation and the Israel Science Foundation\\           
$^{e}$ &  supported by the Italian National Institute for Nuclear Physics (INFN) \\                
$^{f}$ &  supported by the Japanese Ministry of Education, Culture, Sports, Science and Technology 
          (MEXT) and its grants for Scientific Research\\                                          
$^{g}$ &  supported by the Korean Ministry of Education and Korea Science and Engineering          
          Foundation\\                                                                             
$^{h}$ &  supported by the Netherlands Foundation for Research on Matter (FOM)\\                   
$^{i}$ &  supported by the Polish State Committee for Scientific Research, grant no.               
          620/E-77/SPB/DESY/P-03/DZ 117/2003-2005 and grant no. 1P03B07427/2004-2006\\             
$^{j}$ &  partially supported by the German Federal Ministry for Education and Research (BMBF)\\   
$^{k}$ &  supported by RF Presidential grant N 8122.2006.2 for the leading                         
          scientific schools and by the Russian Ministry of Education and Science through its grant
          Research on High Energy Physics\\                                                        
$^{l}$ &  supported by the Spanish Ministry of Education and Science through funds provided by     
          CICYT\\                                                                                  
$^{m}$ &  supported by the Particle Physics and Astronomy Research Council, UK\\                   
$^{n}$ &  supported by the US Department of Energy\\                                               
$^{o}$ &  supported by the US National Science Foundation. Any opinion,                            
findings and conclusions or recommendations expressed in this material                             
are those of the authors and do not necessarily reflect the views of the                           
National Science Foundation.\\                                                                     
$^{p}$ &  supported by the Polish Ministry of Science and Higher Education                         
as a scientific project (2006-2008)\\                                                              
$^{q}$ &  supported by FNRS and its associated funds (IISN and FRIA) and by an Inter-University    
          Attraction Poles Programme subsidised by the Belgian Federal Science Policy Office\\     
$^{r}$ &  supported by the Malaysian Ministry of Science, Technology and                           
Innovation/Akademi Sains Malaysia grant SAGA 66-02-03-0048\\                                       
\end{tabular}                                                                                      

%------------------------------------------------------------------------------
%       Text
%------------------------------------------------------------------------------
\newpage
\pagenumbering{arabic} 
\hyphenation{pa-ra-me-ter con-ta-mi-na-ted sig-ni-fi-cant ge-ne-ra-ted ne-ver-the-less}
\pagestyle{plain}
% ----------------------------------------------------------------------------
%       Introduction
% ----------------------------------------------------------------------------
\section{Introduction}
\label{sec-int}
The use of Bose-Einstein correlations (BEC) in particle physics as 
a method of determininig the size and the shape of the source from
which particles originate was first considered  by Goldhaber et~al. 
\cite{prl:3:181,pr:120:300} in 1959 for $p\bar{p}$ annihilations.
Bose-Einstein correlations originate from the symmetrization of the 
two-particle wave function of identical bosons and lead to an enhancement 
of boson pairs emitted with small relative momenta.
The effect is sensitive to the size of the emitting source.
The studies of BEC for pairs of identical particles have been carried out 
in a large variety of particle interactions. In particular, H1 and ZEUS have
reported results on inclusive charged particle pairs in $e^{\pm}p$ collisions
at HERA~\cite{zpc:75:437,pl:b583:231}.
Recent reviews \mcite{rpp:66:481,hip:15:1,nucl-th/0505019,ibe:weiner,smd:kittel}
summarise the underlying theoretical concepts and experimental results. 
The measurements of the radius of the emission source have been
mostly performed for neutral and charged pions. 
For other bosons, e.g. kaons, the information is scanty.

This paper reports first results on BEC for pairs 
of charged ($K^{\pm}$$K^{\pm}$) and neutral ($K^0_SK^0_S$) kaons in 
deep inelastic scattering (DIS) at HERA.
The measurements are compared with results from $e^+e^-$ interactions 
whose fragmentation properties are expected to be similar to the current
region of DIS~\cite{epj:c11:251}. 
However, proton fragmentation may lead to a significant
difference in the properties of the hadronic final state.

The correlation function for two identical kaons is defined as

\begin{equation}
R(Q_{12})=\frac{P(Q_{12})}{P_0(Q_{12})},
\label{eq:no1}
\end{equation}

where $Q_{12}$ is the four\--momenta difference of the kaons with
four momenta $\textrm{p}_1$ and $\textrm{p}_2$ given as

\begin{equation}
Q_{12}=\sqrt{-(\textrm{p}_1-\textrm{p}_2)^2}=\sqrt{M_{KK}^2-4m_{K}^2},
\label{eq:no2}
\end{equation}

where $M_{KK}$ is the invariant mass of the pair of kaons and $m_{K}$
is the kaon rest mass.
The function $P(Q_{12})$ in Eq.~(\ref{eq:no1}) is the two-particle density:
$P(Q_{12})$ = $(1/N)(dn_{KK}/dQ_{12})$, where $n_{KK}$ is the number
of kaon pairs and $N$ is the number of events. The denominator of Eq.~(\ref{eq:no1}), 
$P_0(Q_{12})$, is the two-particle density in the absence of BEC.

For a static source with a Gaussian density distribution,
the correlation function can be parametrised as follows \cite{pr:120:300}:

\begin{equation}
R(Q_{12})=1+\lambda \exp(-r^2Q_{12}^2).
\label{eq:no3}
\end{equation}

The $\lambda$ parameter gives information about the strength of the BEC. 
For a completely coherent source, $\lambda$ is zero, while for a 
completely incoherent source, $\lambda$ is one \cite{prl:10:84}. 
Contributions from decays of short-lived resonances can further modify 
this parameter. The parameter $r$ is related to the size of the source 
and is called the radius in the  following sections.

% ----------------------------------------------------------------------------
%       Experimental set-up
% ----------------------------------------------------------------------------
\section{Experimental set-up}
\label{sec-exp}
The analysis was performed with data taken by the ZEUS detector between
$1996$ and $2000$ at HERA. The data from $e^{\pm}p$ collisions collected 
in this period with electron\footnote{Here and in the following, the term
,,electron" denotes generically both the electron and the positron.} energy
$E_e=27.5 \gev$ and proton energy $E_p=820\gev$ $(1996\--1997)$ or 
$E_p=920\gev$ \mbox{$(1998\--2000)$} correspond to an integrated luminosity of 
$121 \pbi$.

\Zdetdesc

\Zctddesc

\ZdedxA

\Zcaldesc

The energy of the scattered electron was corrected for energy loss in the 
material between the interaction point and the calorimeter using
the small-angle rear tracking detector \mcite{epj:c21:443, nim:a401:64}
and the presampler \mcite{epj:c21:443, nim:a382:419}.

% ----------------------------------------------------------------------------
%       Event and track selection
% ----------------------------------------------------------------------------
\section{Event and track selection}
\label{sec-evn}
The inclusive neutral current DIS process  $e(k)+p(P)\rightarrow e(k')+X$ 
can be described in terms of the following kinematic variables: $Q^2$,  
the virtuality of the exchanged photon, $x$, the Bjorken 
scaling variable and $y$, the fraction of the lepton energy transferred 
to the proton in the proton rest frame.
They are defined as follows: $Q^2=-q^2=-(k-k')^2$;
$x=Q^2/(2P\cdot q)$; $y=(q\cdot P)/(k\cdot P)$, where $k$, $k'$ and $P$ are
the four-momenta of initial and final scattered electrons and incoming 
proton, respectively. These variables were reconstructed
using the electron method (denoted by the subscript $e$), which requires
measurements of the energy and angle of the scattered electron.

A three-level trigger system \cite{zeus:1993:bluebook} 
was used to select events online. At the third level, electrons with 
energy greater than $4\gev$ and position outside a rectangle defined by 
$|{\rm X}|<12\cm$, $|{\rm Y}|<6\cm$ on the face of the RCAL were accepted.
Data below $Q^2\sim 20\gev^2$ were prescaled to reduce the trigger rate.

The offline selection of DIS events was based on the following
requirements:
\begin{itemize}
\item $|\textrm{Z}_{\rm vtx}|<50\cm$, where $\textrm{Z}_{\rm vtx}$ is the \textrm{Z}
component of the primary-vertex position determined from the tracks. This cut
reduces the background from non-{\it ep} interactions;
\item an identified scattered electron in the CAL with energy $E_e\geq 8.5\gev$;
\item $2~\leq~ Q_e^2~\leq~ 15000\gev^2$;
\item $35<\delta < 60\gev$, where $\delta=\sum E_i(1-\cos \theta_i)$, $E_i$ is
the energy of the $i^{th}$ calorimeter cell and $\theta_i$ is its polar angle 
as viewed from the primary vertex. The sum runs over all CAL cells. This cut reduces 
the background from photoproduction and events with large initial-state 
radiation;
\item $y_e~ \leq~ 0.95$, to remove events with misidentified scattered electrons; 
\item $y_{\rm JB} \geq 0.04$, to remove events with low hadronic activity, where
$y_{\rm JB}$ is the value of $y$ reconstructed using the Jacquet-Blondel method\cite{psfe:391}.
\end{itemize}

Good quality tracks measured in the CTD with high acceptance and resolution 
were selected using the following requirements: transverse momentum 
$p_T>0.15\gev$ and pseudorapidity $|\eta|<1.75$. In addition, 
the tracks were required to pass through more than three CTD superlayers. 

After the above cuts, the data sample contained 25 million events
with at least two good tracks and the average $Q^2$ of the sample was 
$\langle Q^2\rangle=35\gev^2$.
\subsection{Selection of charged kaons}
\begin{figure}[!hp]
\begin{center}
\includegraphics[scale=1]{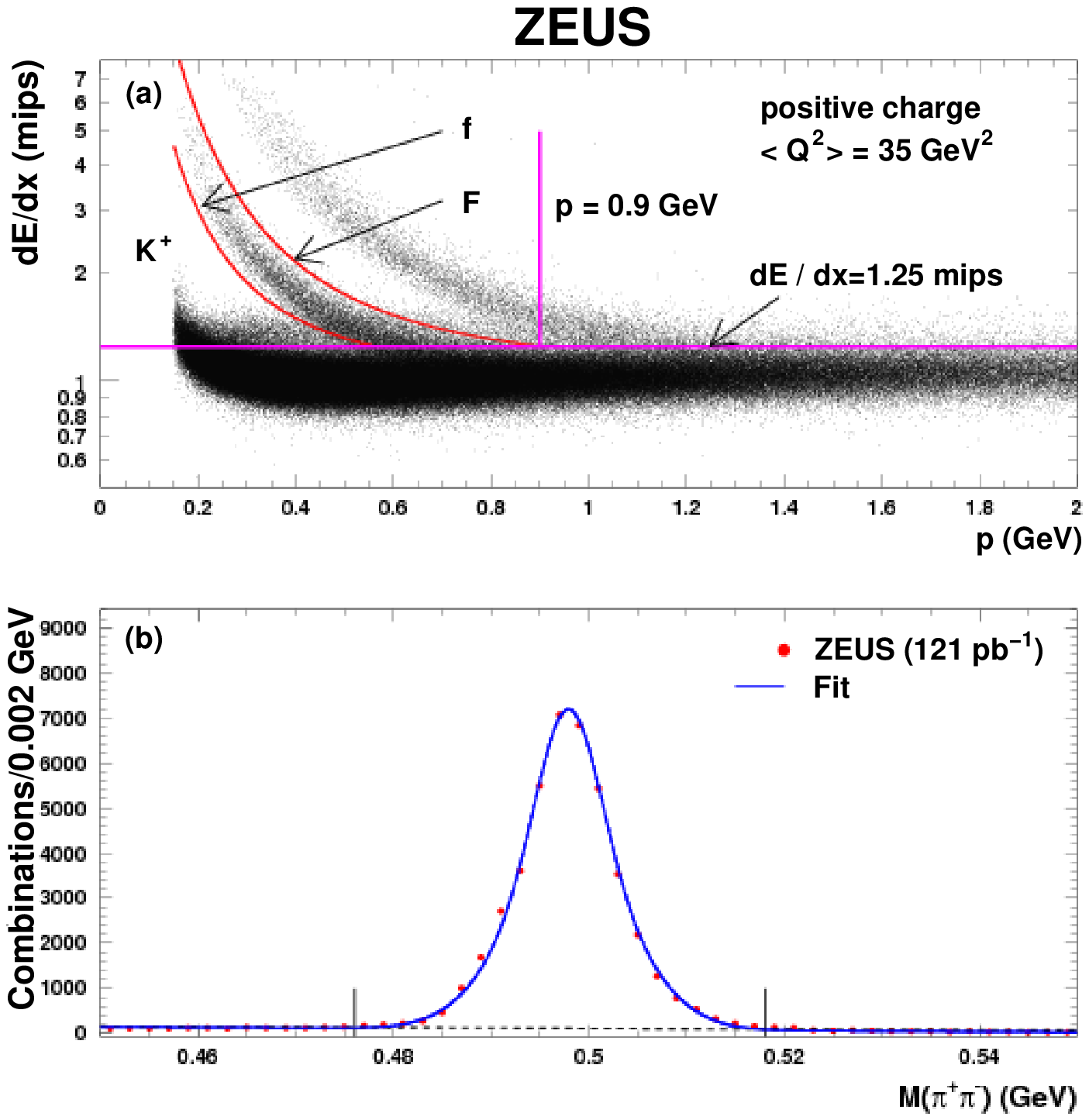}
\end{center}
\caption{(a)~The energy-loss $dE/dx$ as a function of the momentum $p$ for 
tracks with positive charge. The tracks with $f<dE/dx<F$, $dE/dx>1.25$ mips 
and $p<0.9\gev$ were taken as $K^+$.
(b)~The $\pi^+ \pi^-$ invariant-mass distribution of the $K^0_S$ candidates. 
The solid line shows the result of a fit with the sum of two Gaussian functions
and linear background. 
The two Gaussians were fixed to the same
mean value. A mass in agreement with the Particle Data Group value \protect\cite{pdg} 
and a width of $3.2\mev$ ($6.3\mev$)
were obtained from the fit for the central (second) Gaussian. 
The dashed line shows the linear background. The short vertical lines show 
the mass window used to define the signal.
}
\label{fig:dedx+k0signal}
\end{figure}

Charged kaons were selected using the energy-loss measurement, $dE/dx$. 
The analysis used all tracks fitted to the primary vertex
with the exception of the scattered-electron track.
Tracks were selected as described above.
The $dE/dx$ as a function of momentum for positively
charged tracks is shown in Fig.~\ref{fig:dedx+k0signal}a.
The curves indicate the region used for identification of 
positively charged kaons. Kaons were selected by requiring $f<dE/dx<F$,
where $f$ and $F$ are functions of the track momentum, $p$, 
motivated by the Bethe-Bloch equation. For positive kaons
$f=0.08/p^2+1.0$, $F=0.17/p^2+1.03$ mips and for negative kaons 
$f=0.08/p^2+1.0$ and $F=0.18/p^2+1.03$ mips (with $p$ in \gev).
The slight difference arises from the different response of the CTD
to positive and negative tracks. The kaons were identified for $p<0.9\gev$
and $dE/dx>1.25$ mips.

The kaon identification efficiency for $p_t>0.15\gev$, $|\eta|<1.5$ and
$p<0.9\gev$ was $61\%$, with a purity of $90\%$. The resulting data 
sample contained 55522 $K^+K^+$ or $K^-K^-$ pairs.

\subsection{Neutral kaon selection}
The $K^0_S$ mesons were identified using the charged\--decay channel 
$K_S^0\rightarrow \pi^+\pi^-$ with a similar selection as in a previous 
publication \cite{pl:b591:7}. 
The pion tracks were required to originate from secondary vertices. 
Assigning the pion mass to both tracks, the invariant 
mass $M(\pi^+\pi^-)$ was calculated and the candidate was accepted if the mass
was within $\pm ~20\mev$ of the nominal PDG~\cite{pdg} $K^0_S$ mass.
To eliminate tracks from photon conversions and $\Lambda/\bar{\Lambda}$ 
contamination, the electron, pion and proton masses were assigned to 
tracks and the following cuts were used: $M(e^+e^-)>80\mev$ and 
$M(\pi p)>1121\mev$.

The following additional requirements were applied to the selected $K^0_S$ candidates:
\begin{itemize}
\item 
$2<L_d<30\cm$, where $L_d$ is the decay length of the 
$K^0_S$ candidate;
\item
$\Delta \textrm{Z}<0.8\cm$, where $\Delta\textrm{Z}$ is the projection 
on the $\textrm{Z}$ axis of the vector defined by the primary interaction 
point and the point of closest approach of the $K^0_S$ candidate;
\item
$\alpha_{\rm XY}<8^{\circ}$, where $\alpha_{\rm XY}$ is the (collinearity) 
angle between the candidate $K^0_S$ momentum vector and the vector defined 
by the interaction point and the $K^0_S$ decay vertex in the ${\rm XY}$ plane;
\item
$p_{t}^{\rm PA}>0.11 \gev$,  where the Podolanski-Armenteros 
variable $p_{t}^{\rm PA}$ is the projection of the candidate pion momentum 
onto a plane perpendicular to the $K^0_S$ momentum direction~\cite{pm:45:13}.
\end{itemize}

The total number of $K^0_S$ candidates was $725505$.
After all cuts, the selected data sample contained 19494 $K^0_SK^0_S$ pairs 
and 400 triples. Each combination of two particles was included in the analysis.
Figure~\ref{fig:dedx+k0signal}b shows the $\pi^+\pi^-$ invariant mass 
distribution after the $K^0_S$-pair selection and a fit to the signal plus 
linear background, which resulted in an estimated background of $1.4\%$.
% ----------------------------------------------------------------------------
%       Monte Carlo simulations
% ----------------------------------------------------------------------------
\section{Monte Carlo simulation}
\label{sec-mc}
Inclusive DIS events with $Q^2>2\gev^2$  were 
generated without BEC using the {\sc Ariadne} 4.10 Monte Carlo (MC)
model \cite{cpc:71:15} interfaced with {\sc Heracles} 4.6.1 \cite{cpc:69:155}
via the {\sc Djangoh} 1.1 program \mcite{cpc:81:381, heracles:djangoh:web} 
in order to incorporate first-order electroweak corrections.
The Lund string model \cite{pr:97:31} was used for the description of 
hadronisation, as implemented in the {\sc Jetset} 7.4 \cite{cpc:82:74} program.

The generated events were passed through a full simulation of the detector
using the {\sc Geant} 3.13 program \cite{cern-rep:1987} and reconstructed and analyzed
in the same way as the data. The {\sc Ariadne} MC sample corresponds to a similar
integrated luminosity as that of the data.
% ----------------------------------------------------------------------------
%       Analysis procedure 
% ----------------------------------------------------------------------------
\section{Extraction of BEC parameters}
\label{sec-anal}
The main difficulty in measuring BEC is in the construction of a reference
sample which should be as close as possible to the analyzed data in all aspects 
but free from the Bose-Einstein effect.
The obvious reference sample provided by unlike-sign charged kaon pairs
cannot be used due to the strong signal of the $\phi^0(1020)\rightarrow K^+K^-$
decay at low values of $Q_{12}$.

A reference sample can  be derived from a Monte Carlo simulation without BEC.
In this so-called single-ratio method, the correlation function $R(Q_{12})$
is defined as: $R^S(Q_{12}) = P(Q_{12})^{\rm data} / P(Q_{12})^{\rm MC,noBEC}$, 
where $P(Q_{12})^{\rm data}$  is the normalized two-particle density distribution 
for the data and $P(Q_{12})^{\rm MC,noBEC}$ is the corresponding distribution
obtained for MC without BEC.
However, this approach requires a correct simulation of the physics 
processes in the absence of BEC, as well as a good description of the 
detector effects.

In another approach, a reference sample can be obtained using an event-mixing 
procedure where two kaons from different events are combined. 
This method, which is used in this analysis, is less sensitive to 
imperfections in the MC simulation. To correct for other correlations 
lost in the event-mixing procedure, the two-particle correlation 
function $R(Q_{12})$ was calculated using the double-ratio method:

\begin{equation} 
R(Q_{12}) = \frac{P(Q_{12})^{\rm data}}{P_{\rm mix}(Q_{12})^{\rm data}}  
\Bigg/ \frac{P(Q_{12})^{\rm MC, noBEC}}{P_{\rm mix}(Q_{12})^{\rm MC, noBEC}},
\label{eq:no4} 
\end{equation}

where $P_{\rm mix}(Q_{12})^{\rm data}$ is the two-particle density 
constructed from pairs of kaons coming from different events
and $P_{\rm mix}(Q_{12})^{\rm MC,noBEC}$ is obtained in a similar way for 
MC events.
The double-ratio method was used for the main analysis and the single-ratio
method only to estimate systematic uncertainties.

To fit the correlation function $R(Q_{12})$ defined by Eq.~(\ref{eq:no4}), 
the modified Goldhaber parametrisation (Eq.~(\ref{eq:no3})) multiplied by
an empirical term $1+\beta Q_{12}$, which accounts for the presence of 
possible long-range two-particle correlations for high $Q_{12}$, is often used:

\begin{equation} 
R(Q_{12}) = \alpha (1+\lambda e^{-Q_{12}^2r^2})(1+\beta Q_{12}).
\label{eq:no5} 
\end{equation}

Such correlations are imposed for example by energy and charge conservation, 
phase-space constraints or strangeness compensation.
In this analysis, the $\beta$ parameter was found to be zero within errors
and its possible deviation from zero was included in the systematic uncertainties.
% ----------------------------------------------------------------------------
%       Systematic
% ----------------------------------------------------------------------------
\section{Systematic uncertainties}
Systematic uncertainties on the fitted $\lambda$ and $r$ parameters 
arise from event and track selection, the modeling of $dE/dx$, 
the fitting procedure and the construction of the reference sample. 
They were calculated from the deviation of the fit parameters from their 
nominal values after changing the analysis cuts or procedures.
For charged kaons, a bias originating from the contamination of the experimental 
kaon sample by pion, proton and antiproton is expected.
For neutral kaons, a bias can be introduced by the contamination of the sample 
by $K^0_SK^0_S$ pairs from the decay of $f_0(980)$ resonance. These effects are
discussed in the next sections.

The following systematic studies were carried out for the charged kaon sample.
The resulting changes for $\lambda$ and $r$ are given in parentheses
as $[\Delta\lambda, \Delta r]$:
\begin{itemize}
\item 
the correlation function was calculated using the single-ratio method
$[-0.01,-0.03]$;
\item
the fit was repeated for different lower and upper limits of $Q_{12}$.
In addition the binning in $Q_{12}$ was modified
$[^{+0.04}_{-0.04}, ^{+0.09}_{-0.04}]$; 
\item
the momentum range was reduced to $p<0.7\gev$ for both the double 
$[+0.06, +0.06]$ and single $[+0.01, -0.01]$ ratio methods; 
\item
the definition of the kaon band was varied for the double-ratio method
$[-0.02, +0.05]$ and the single-ratio method $[-0.04, -0.02]$;
\item
the track quality cuts were changed within their resolutions $[-0.02, +0.04]$;
\item
the DIS selection cuts on $E_{e}$, $y_{e}$, $y_{\rm JB}$, $\delta$ 
were varied within the resolutions $[^{+0.02}_{-0.02}, ^{+0.02}_{-0.02}]$;
\item
the influence of pion, proton and antiproton contamination was checked by 
tightening and relaxing the kaon selection criteria. The purity was raised 
to $99\%$ $[+0.05, +0.07]$ and was brought down to $80\%$ $[-0.04, -0.05]$.
\end{itemize}

The following systematic uncertainties for neutral kaons were considered:

\begin{itemize}
\item 
the correlation function was calculated using the single-ratio method
$[+0.10, +0.02]$;
\item 
the fit was repeated for different lower and upper limits of $Q_{12}$. In 
addition, the binning in $Q_{12}$ was modified $[^{+0.04}_{-0.04}, 
^{+0.09}_{-0.04}]$; 
\item 
the cuts on the $K^0_S$ momentum, $\alpha_{\rm XY}$, $\Delta\textrm{Z}$, 
$M(\pi^+\pi^-)$ were varied $[-0.08, -0.08]$;
\item 
the cut on $p_{t}^{\rm PA}$ was changed to $0.12\gev$ $[+0.04, +0.02]$;
\item 
the full Eq.~(\ref{eq:no5}) was used to check the sensitivity of the fit 
to possible long-range correlations $[+0.11, +0.03]$; 
\item 
the track quality cuts were changed within the resolution $[+0.15, +0.02]$;
\item 
the cut to remove the $e^+e^-$ background was changed to $M(e^+e^-)>50 \mev$
$[+0.08, +0.03]$;
\item 
the DIS selection cuts on $E_{e}$, $y_{e}$, $y_{\rm JB}$, $\delta$ were 
varied within resolutions $[^{+0.02}_{-0.02},^{+0.02}_{-0.02}]$.
\end{itemize}

The contributions from the different groups of systematic uncertainties
in the parameters $\lambda$ and $r$ were added in quadrature separately 
for positive and negative variations.
The overall systematic uncertainties in $\lambda$ and 
$r$ for $K^{\pm}K^{\pm}$ and $K^0_SK^0_S$ are presented in Table~\ref{tab1}.     
% ----------------------------------------------------------------------------
%       Results
% ----------------------------------------------------------------------------
\section{Results}
\label{sec-res}
\subsection{Correlations in $K^{\pm}K^{\pm}$ pairs}
\begin{figure}[!hp]
\begin{center}
\includegraphics[scale=0.75]{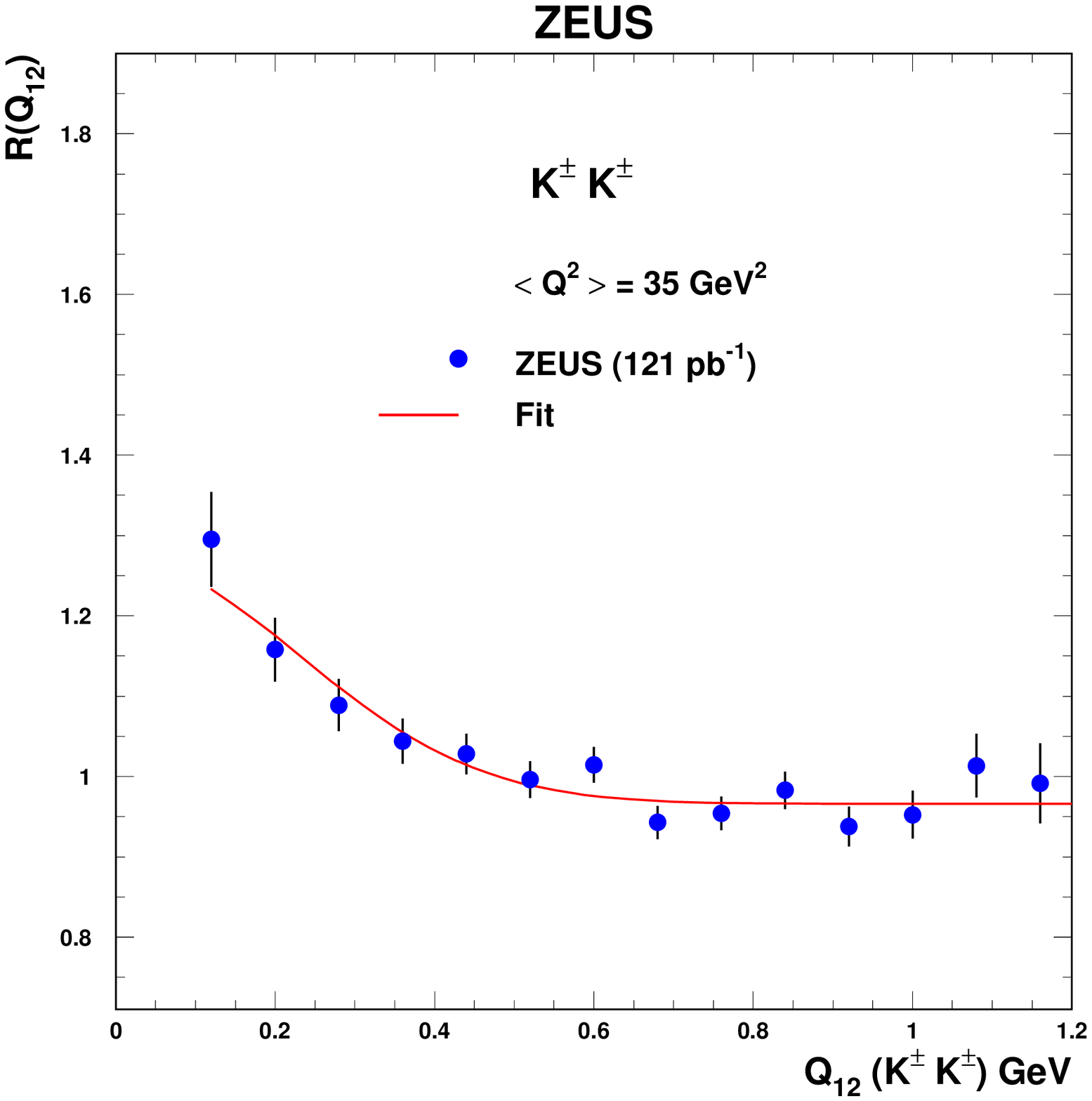}
\end{center}
\caption{
The two-particle correlation function for charged kaons with a fit to 
Eq.~(\ref{eq:no5}), with $\beta=0$. The error bars represent the statistical 
uncertainties.}
\label{fig:k+bec}
\end{figure}

Figure \ref{fig:k+bec} shows the measured two-particle correlation
function for $K^{\pm}K^{\pm}$ pairs. The results obtained by the fit
function given by Eq.~(\ref{eq:no5}) with $\beta=0$ are $r=0.57\pm0.09$ 
fm and $\lambda=0.31 \pm 0.06$. The $90\%$ purity of the kaon selection 
introduces in the kaon pair sample an $18\%$ admixture of unlike particles
pairs which are not correlated. The measured $\lambda$ is therefore expected 
to be underestimated by $18\%$. The systematic checks involving purity 
confirm this expectation. 
After the correction for purity, the result is $\lambda=0.37 \pm 0.07^{+0.09}_{-0.08}$.
The value of the radius is not affected by this correction and the
result is $r=0.57\pm0.09^{+0.15}_{-0.08}$ fm. The corrected parameters of BEC 
correlations for $K^{\pm}K^{\pm}$ pairs are presented in Table~\ref{tab1}. 
The radius value is consistent with that for charged particles:
$r=0.54\pm0.03^{+0.03}_{-0.02}$ fm for H1~\cite{zpc:75:437} and 
$r=0.666\pm0.009^{+0.022}_{-0.036}$ fm for ZEUS~\cite{pl:b583:231}. 
The obtained value of $\lambda$ for kaons is in agreement with the H1 result
for charged particles $\lambda=0.32\pm0.02 \pm 0.06$, although somewhat 
smaller than the ZEUS result $\lambda=0.475\pm0.007^{+0.011}_{-0.003}$.  
\subsection{Correlations in $K^0_SK^0_S$ pairs}
\begin{figure}[!hp]
\begin{center}
\includegraphics[scale=0.75]{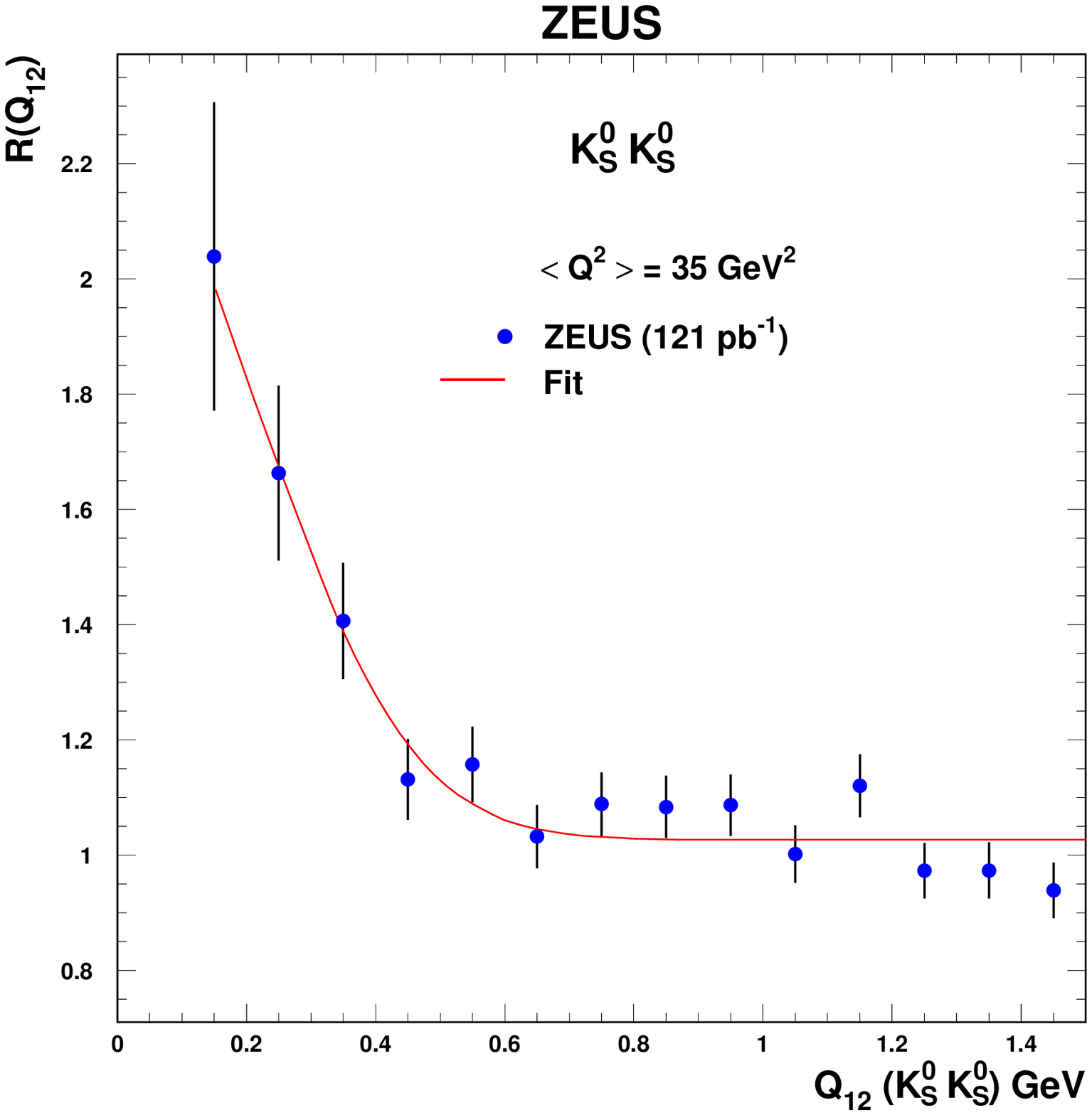}
\end{center}
\caption{
The two-particle correlation function for neutral kaons with a fit to  
Eq.~(\ref{eq:no5}), with $\beta=0$. The error bars represent the statistical 
uncertainties.}
\label{fig:k0bec}
\end{figure}

The $K^0_SK^0_S$ pairs may originate not only from $K^0K^0$, 
$\bar{K^0}\bar{K^0}$ (strangeness $S=\pm 2$) states but also 
from the $K^0\bar{K^0}$ (strangeness $S=0$) system, which may come 
from the decay of resonances. It has been shown~\mcite{pl:b219:474,prl:69:3700}
that a Bose-Einstein-like enhancement is nevertheless expected in  
the $Q_{12}$ distribution of the $K^0_SK^0_S$ pairs, even when their
origin is from the $K^0\bar{K^0}$ system.
According to the MC simulation of DIS events, about $75\%$ of low-$Q_{12}$ 
$K^0_SK^0_S$ pairs come from $K^0\bar{K^0}$. Similar to the case of charged 
kaons, Eq.~(\ref{eq:no5}) with $\beta=0$ was used to fit the correlation 
function calculated for $K^0_SK^0_S$  pairs. The so-called raw results 
are $\lambda=1.16\pm 0.29^{+0.28}_{-0.08}$ and $r=0.61\pm 0.08^{+0.07}_{-0.08}$ fm.
The results are presented in Table~\ref{tab1} and Fig.~\ref{fig:k0bec}.
The measured radii for $K^{\pm}K^{\pm}$ and $K^0_SK^0_S$ are close to each 
other. 

\begin{figure}[!hp]
\begin{center}
\includegraphics[scale=0.75]{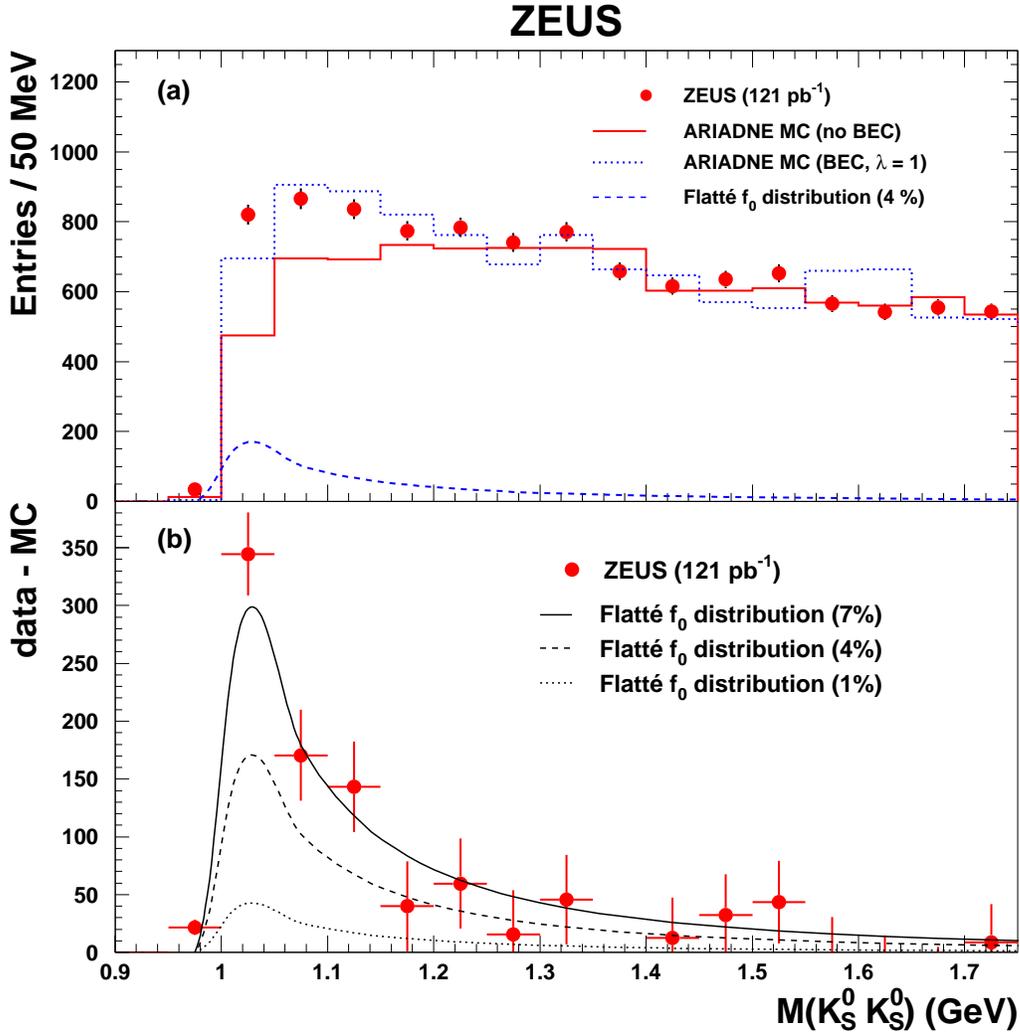}
\end{center}
\caption{(a)~$K^0_SK^0_S$ invariant mass for data (points) and Monte Carlo
with BEC (dotted histogram) and without BEC (full-line histogram).
The dashed line shows the Flatt\'e function used to describe
the $f_0$ resonance.
(b)~The difference between data and MC without BEC compared
to different $f_0$ contributions.
}
\label{fig:k0.flatte}
\end{figure}

For $K^0_SK^0_S$ the fit does not take into account a possible contamination 
from the scalar $f_0(980)$ resonance decaying below the $K\bar{K}$ threshold. The decay 
channel $f_0(980)\rightarrow K^0\bar{K^0}$ is not included in the standard MC, 
therefore the $f_0(980)$ 
contribution may distort the strength of the Bose-Einstein effect. 
In order to investivate this, a MC sample with BEC at maximum strength
was generated. A comparison of this MC with the data showed an excess for 
$Q_{12}<0.6\gev$, which may indicate a possible $f_0$ contribution. 
The same excess is also visible in the $M_{KK}$ distribution, as 
illustrated in Fig.~\ref{fig:k0.flatte}a.

The expected contribution from $f_0$ can be approximated by 
a modified Breit-Wigner function as proposed by Flatt\'e~\mcite{pl:b63:224,pl:b63:228,pl:b611:66}:

\begin{equation}
\frac{d\sigma}{dM_{KK}} = N_F \cdot \frac{m_0^2 \cdot \Gamma_{KK}}{(m_0^2-M_{KK}^2)^2+(m_0\cdot(\Gamma_{\pi\pi}+\Gamma_{KK}))^2},
\label{eq:flatte}
\end{equation}

where $m_0=0.954 \gev$ is the mass of $f_0(980)$ and the widths $\Gamma_{\pi\pi}$ 
and $\Gamma_{KK}$ are related to the coupling constants $g_{\pi}=0.11$ and 
$g_{K}=0.423$ by $\Gamma_{\pi\pi}=g_{\pi}\sqrt{M^2_{KK}/4-m^2_{\pi}}$ and
$\Gamma_{KK}=g_{K}\sqrt{M^2_{KK}/4-m^2_{K}}$.
The normalization factor, $N_F$, was adjusted to give the total number of 
$f_0(980)\rightarrow K^0_SK^0_S$ decays for $M_{KK}<1.8 \gev$. 
Fig.~\ref{fig:k0.flatte}a shows the distribution of ${d\sigma}/{dM_{KK}}$
according to Eq.~(\ref{eq:flatte}).
Figure~\ref{fig:k0.flatte}b shows the $M_{KK}$ distribution in the data after 
subtraction of the standard MC without BEC together with different 
contributions of the $f_0$ resonance.
An $f_0$ contribution of about $7\%$ is sufficient to account for 
the data, therefore, the present analysis cannot distinguish the Bose-Einstein 
effect from the effect of an $f_0$ contribution. 
Assuming that both BEC and $f_0$ are present, the amount of $f_0$ was estimated
from the shape of the $R(Q_{12})$ distribution (Eq.~(\ref{eq:no4})).
For this purpose, fits of the correlation function were 
carried out with different percentages, $c_{f_0}$, of $f_0$ subtracted
from the data. A shallow minimum of the $\chi^2(c_{f_0})$ value for 
$c_{f_0}=4\%$ was found, with one-sigma limits of $c_{f_0}=1\%$ and 
$c_{f_0}=7\%$. The values of $\lambda$ and $r$ at 
$c_{f_0}=4\%$ were taken as the most probable result and their uncertainties 
were calculated from the one sigma limits of $c_{f_0}$. 

The results corrected for the $f_0$ contamination are included in 
Table~\ref{tab1}. After adding the uncertainties coming from the $f_0$ subtraction 
to systematics, the corrected results are $\lambda=0.70\pm 0.19^{+0.47}_{-0.53}$ 
and $r=0.63\pm 0.09^{+0.11}_{-0.08}$ fm. 
The uncertainty on a possible $f_0$ admixture leads to a large systematic 
uncertainty on $\lambda$. The radius $r$ is less sensitive to the $f_0$ admixture.
The radii for $K^0_SK^0_S$ and $K^{\pm}K^{\pm}$ are very similar. 
%     
% ----------------------------------------------------------------------------
%       Comparison with LEP
% ----------------------------------------------------------------------------
%
\subsection{Comparison with $e^+e^-$ interactions}
\label{sec-comparison}
\begin{figure}[!hp]
\begin{center}
\includegraphics[scale=0.75]{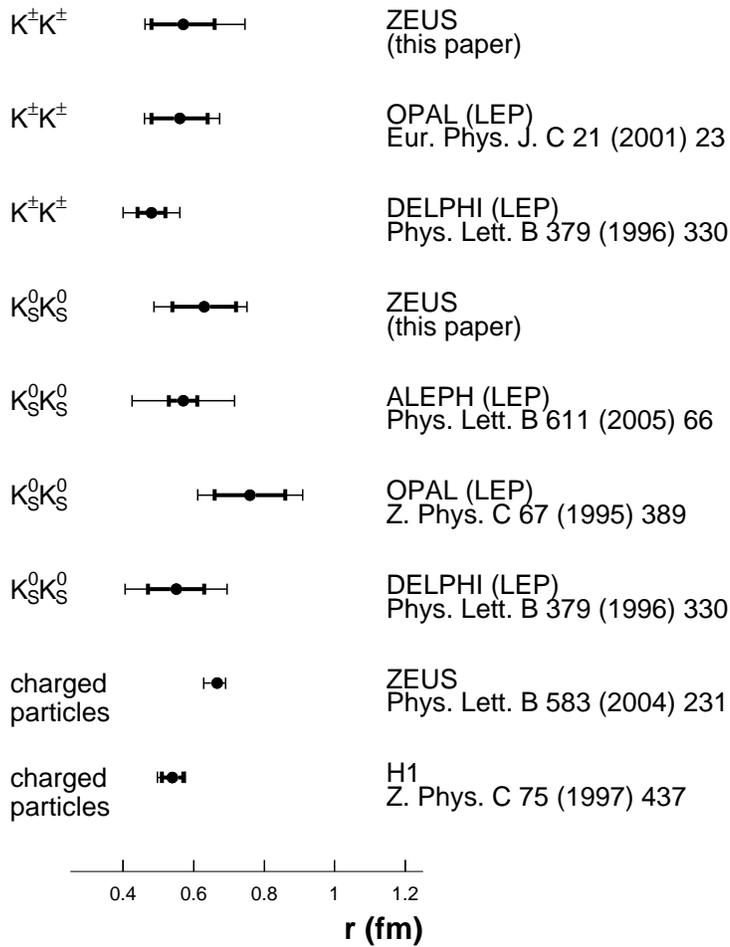}
\end{center}
\caption{
Comparison of DIS and LEP results for r obtained from
Bose-Einstein correlation studies of charged and neutral 
kaons and unidentified charged particles. The DIS result for 
neutral kaons is corrected for the $f_0$ contribution. 
}
\label{fig:k0k+radius}
\end{figure}

Figure~\ref{fig:k0k+radius} shows the comparison for $r$ between DIS and 
$e^+e^-$ annihilation results at LEP \cite{pl:b379:330,epj:c21:23,zpc:67:389,pl:b611:66}
for both charged and neutral kaons. The radius value obtained in DIS agrees 
with the measurements from LEP. The same figure also presents DIS results 
for unidentified charged particles. The kaon results agree with those for 
charged particles within systematic uncertainties. The $\lambda$ value for 
charged kaons in DIS is somewhat smaller than that in $e^+e^-$ collisions,
which may be related to the fact that the DIS data mostly populate the proton 
fragmentation region in the Breit frame. 
%                                                                  
% ----------------------------------------------------------------------------
%       Conclusions
% ----------------------------------------------------------------------------
\section{Conclusions}
\label{sec-con}
Bose-Einstein correlations have been measured for pairs of charged
and neutral kaons in deep inelastic scattering at HERA using the ZEUS detector.
The values of the radius for charged and neutral kaons agree within systematic 
uncertainties. They are also consistent with the measurements for charged 
particles in DIS and kaons in $e^+e^-$ collisions at LEP.
The $f_0(980)\rightarrow K^0_SK^0_S$ decay can significantly affect the 
$\lambda$ parameter for $K^0_SK^0_S$ correlations.

\section{Acknowledgments}
We wish to thank the DESY Directorate for their strong support and 
encouragement. The remarkable achievements of the HERA machine group 
were essential for the successful completion of this work and are greatly
appreciated. We are grateful for the support of the DESY computing and
network services. The design, construction and installation of the ZEUS
detector have been made possible owing to the ingenuity and effort of many
people from DESY and home institutes who are not listed as authors.

%------------------------------------------------------------------------------
%       Bibliography
%------------------------------------------------------------------------------
\providecommand{\etal}{et al.\xspace}
\providecommand{\coll}{Coll.\xspace}
\catcode`\@=11
\def\@bibitem#1{%
\ifmc@bstsupport
  \mc@iftail{#1}%
    {;\newline\ignorespaces}%
    {\ifmc@first\else.\fi\orig@bibitem{#1}}
  \mc@firstfalse
\else
  \mc@iftail{#1}%
    {\ignorespaces}%
    {\orig@bibitem{#1}}%
\fi}%
\catcode`\@=12

%------------------------------------------------------------------------------
%       Tables
%------------------------------------------------------------------------------
\begin{table}[!hp]
\begin{center}
\begin{tabular}{|c|c|c|}
\hline
  & \tablehead{1}{$\lambda$}
  & \tablehead{1}{$r$ [fm]}\\
\hline
&&\\
$K^{\pm}K^{\pm}$ (corrected) & $0.37 \pm 0.07^{~+0.09}_{~-0.08}$& $0.57\pm0.09^{~+0.15}_{~-0.08}$ \\
&&\\
$K^0_SK^0_S$ (raw) & $1.16\pm0.29^{~+0.28}_{~-0.08}$& $0.61\pm0.08^{~+0.07}_{~-0.08}$ \\
&&\\
$K^0_SK^0_S$ (corrected) & $0.70\pm0.19^{~+0.28+0.38}_{~-0.08-0.52}$& $0.63\pm0.09^{~+0.07+0.09}_{~-0.08-0.02}$ \\
&&\\
\hline
\end{tabular}
\caption{
The radius $r$ and the correlation strength $\lambda$ for charged and
neutral kaons extracted from fitting the Goldhaber parametrization
(Eq.~(\ref{eq:no5}) with $\beta =0$) to the Bose-Einstein correlation function.
The first uncertainties are statistical and the second systematic.   
The raw results for $K^0_SK^0_S$ pairs correspond to those shown in 
Fig.~\ref{fig:k0bec}; the corrected results were obtained after 
subtraction of the $f_0$ contribution from the data. For the corrected values 
the third uncertanty comes from the $f_0$ subtraction. 
}
\label{tab1}
\end{center}
\end{table}
\vfill

%------------------------------------------------------------------------------
%       Figures
%------------------------------------------------------------------------------
\vfill
%%%%%%%%%%%%%%%%%%%%%%%%%%%%%%%%%%%%%%
\vfill
%%%%%%%%%%%%%%%%%%%%%%%%%%%%%%%%%%%%%%
\vfill
%%%%%%%%%%%%%%%%%%%%%%%%%%%%%%%%%%%%%%
\vfill
%%%%%%%%%%%%%%%%%%%%%%%%%%%%%%%%%%%%%%
\vfill
%%%%%%%%%%%%%%%%%%%%%%%%%%%%%%%%%%%%%%

\end{document}